\documentclass[prb,aps,twocolumn]{revtex4}
\usepackage{bm}
\usepackage{amsmath,amssymb}
\usepackage{color,multirow}
\usepackage{graphicx}
\begin{document}
\title{Josephson current in a normal-metal nanowire coupled to superconductor/ferromagnet/superconductor junction}

\author{Hiromi Ebisu$^{1}$, Bo Lu$^{1}$, Katsuhisa Taguchi$^{1,2}$,
Alexander A. Golubov$^{3,4}$, Yukio Tanaka$^{1,4}$}
\affiliation{$^1$~Department of Applied Physics, Nagoya University, Nagoya 464-8603, Japan\\
$^2$~Department of Physics, Hong Kong University of Science and Technology, Clear Water Bay, Hong Kong, China\\
$^3$~Faculty of Science and Technology and MESA+ Institute for Nanotechnology, University of Twente, 7500 AE Enschede, The Netherlands\\
$^4$~Moscow Institute of Physics and Technology, Dolgoprudny, Moscow 141700, Russia
}
\date{\today}

\begin{abstract}

We consider superconducting nanowire proximity coupled to
superconductor / ferromagnet / superconductor junction,
where the magnetization penetrates into
superconducting segment in nanowire  decaying as $\sim\exp[-\frac{\mid n \mid}{\xi}]$, where  $n$ is the site index and the $\xi$ is the decay length.
We tune chemical potential and spin-orbit coupling so that
topological superconducting regime hosting Majorana fermion
is realized for long $\xi$.
We find that when $\xi$ becomes shorter, zero energy state at the interface
between superconductor and ferromagnet splits into two states at
nonzero energy.
Accordingly, the behavior of Josephson current is drastically changed due to
this ``zero mode-non-zero mode crossover". By tuning the model parameters, we
find an almost second-harmonic current-phase relation,
$\sin2\varphi$, where $\varphi$ is the phase difference of the
junction. Based on the analysis of
Andreev bound state (ABS), we clarify that the current-phase relation is determined by coupling of the states within the energy gap.
We find that the emergence of  crossing points of ABS is a key ingredient to
generate  $\sin2\varphi$ dependence in the current-phase relation.
We further study both the energy and $\varphi$ dependence of pair amplitudes in the ferromagnetic region.
For large $\xi$, odd-frequency spin-triplet $s$-wave  component is
dominant. The magnitude of the odd-frequency pair amplitude is enhanced
at the energy level of ABS.

\end{abstract}
\pacs{pacs}
\maketitle
\thispagestyle{empty}
\section{Introduction}
It is known that
a number of remarkable physical  phenomena occur in superconductor/ferromagnet (S/F) hybrid structures.
\cite{Golubov_RMP,buzdin_rmp,Efetov2} First one is the generation of $\pi $%
-state \cite{Bulaevskii,buzdin_jetp} in S/F/S junctions. Since the exchange
coupling and spin-singlet Cooper pair are competing each other, spin-singlet
pairs in ferromagnet have a spatial oscillation with changing
sign in the presence of
the exchange coupling. \cite{ff,lo_jetp,ryazanov,kontos,robinson_prl}
Next one is the dominant second-harmonic in the current-phase relation of
Josephson current, $\sin 2\varphi $, where $\varphi $ is the phase difference across the junction.
It is known that $\sin 2\varphi $ dependence \cite{Tanaka1997}
appears near the $0$-$\pi $
transition point.\cite{Sellier,Robinson2007}
The third one is the generation of the odd-frequency pairing in the F region
by proximity effect in
S/F hybrid systems.\cite{Efetov1,Efetov2,volkov}
%
There have been many theoretical \cite{eschrig_pt,Eschrig2003}
and experimental \cite{Keizer,Sosnin,Birge,Robinson,Sprungmann,Anwar} works
about proximity effect via odd-frequency pairing in S/F junctions.
The fourth one is the so called inverse proximity effect where
magnetization penetrates into a superconductor.\cite%
{Bergeret2004,Bergeret2005,Linder2009,Xia2009} The electronic property and
pairing symmetry near the S/F interface is drastically changed by this
effect.

Independently of research directions mentioned above, study of nanowire on the
surface of
superconductor in the presence of applied Zeeman magnetic field has recently become a
hot topic in condensed matter physics.\cite{lutchyn10,oreg10} Due to
the strong spin-orbit coupling (SOC) in nanowire, topological superconducting
state is generated. Then, by the bulk-edge correspondence, superconducting
nanowire hosts Majorana fermion (MF) as the end state,
\cite{Fujimoto09,lutchyn10,oreg10}
which is one of important factors
to realize quantum computation. \cite{Kitaev01,alicea12,alicea11,Nayak}
It has also been reported that a chain of ferromagnetic atoms on a
superconductor forms a topologically non-trivial state,
where the ABS within the superconducting gap is localized
around the edge as a MF.\cite{Yazdani2013,yazdaniex,
Ojanen2014,Pientka2014,Braunecker,Klinovaja2013,Pawlak,Hoffman,Meng}
The common feature in these one-dimensional topological superconducting
systems is that both pair potential and magnetization coexist in all sites of
the nanowire.


Up to now, although inverse proximity effect and topological superconductivity
have been studied independently, they have not been studied simultaneously.
It is a challenging issue to clarify a new effect
 where both effects coexist in the same model.
If we consider proximity coupled nanowire on the S/F/S junction,
we can divide the nanowire into the three segments; left superconductor, middle ferromagnet and right superconductor.
Thus, it is possible to design effective one-dimensional S/F/S junctions in
nanowire. We consider the situation
where the ferromagnetic order of F depends on the
position in the S/F/S junction.
Besides, to discuss topological superconductivity,
we consider Rashba-type SOC in nanowire. If
$\xi$ which represents the penetration length of ferromagnetic order is long, zero energy state is generated at the S/F interface.
On the other hand, if $\xi $ is short, we can expect that this
zero energy state splits into two \cite{Yu,Shiba,Rusinov} around the magnetic impurity.
Thus, proximity coupled nanowire on the S/F/S junction is interesting since
we can study both inverse proximity effect and topological superconductivity in the same model.
If we tune $\xi $ as a parameter, the present model has a unique feature
to study a new-type of inverse proximity effect including topological
superconductivity.

In this paper, we study electronic spectra and resulting Josephson current
in this nanowire S/F/S model. First, we calculate local density of states
(LDOS) of isolated left side superconducting segment where magnetization
penetrates from the right edge proportional to $\exp [\frac{n}{\xi }_{L}]$
with site index $n<0$ and the decay length $\xi _{L}$. We clarify that if $%
\xi _{L}$ exceeds a certain value, LDOS has a clear zero energy peak (ZEP)
due to the zero mode at the edge. On the other hand, when $\xi _{L}$ becomes
shorter, LDOS has a peak splitting.
We call this effect as \textquotedblleft zero
mode-non-zero mode crossover". Throughout this paper, we introduce zero mode and non-zero mode to distinguish zero energy state localized at the interface and splitted state.
We study the Josephson current in the S/F/S
nanowire junction by changing the decay length of ferromagnetic order into
the left (right) side superconductor, $\xi _{L}$ ($\xi _{R}$), and chemical potential of
ferromagnet. It is seen that the behavior of Josephson current is quite
different when the decay length becomes shorter due to the
\textquotedblleft zero mode-non-zero mode crossover".

\par
Especially when this non-zero modes \cite{Yu,Shiba,Rusinov}
are localized at
the boundary between superconductor and ferromagnet, we find an anomalous
current-phase relation which can be roughly expressed as $\sin 2\varphi $.
In order to understand the physical origin of the
current-phase relation more clearly, we
calculate the ABS and provide an argument that the
emergence of crossing points of ABS is a key ingredient
to produce current-phase relation $\sin 2\varphi $.
We find that the emergence of
crossing points of ABS is a key ingredient to generate
$\sin 2\varphi $ dependence in current-phase relation.
This mechanism is  distinct from
preexisting case in S/F/S junctions.
\cite{Golubov_RMP,Tanaka1997,Sellier,Robinson2007}
We also calculate pair
amplitude decomposing into odd-frequency part and even-frequency one.
We focus on $s$-wave component of odd-frequency spin-triplet even-parity (OTE)
pairing 
and show their dependence on energy and phase difference $\varphi $.
For large magnitude of $\xi _{L(R)}$, equal spin OTE pair amplitude becomes
dominant and enhanced at the energy level of the ABS.

This paper is organized as follows: In Sec. \ref{sec2}, we provide the formulation to
calculate LDOS and Josephson current using
recursive Green's function technique. In Sec. \ref{sec3a}, we calculate LDOS on the
edge of isolated left side superconducting segment with decaying
ferromagnetic order from the right edge. In Sec. \ref{sec3b}, we analyze the
Josephson current in S/F/S junction changing decay length of ferromagnetic
order parameter and chemical potential of ferromagnet. In Sec. \ref{sec3c}, we
calculate ABS and discuss the relevance to current-phase relation in Sec. \ref{sec3b}.
In Sec. \ref{sec3d}, we calculate pair amplitudes to shed light on these results
from different angles. Finally, we summarize our results in Sec. \ref{sec4}.

\section{Formulation}\label{sec2}

In this section we provide a formulation to calculate LDOS and Josephson
current using recursive Green's function. The method of calculating ABS is
also included in this section.

\begin{figure}[h]
\begin{center}
\includegraphics[width=80mm]{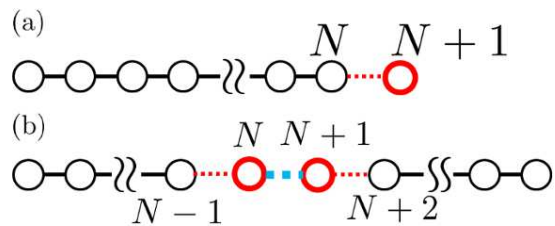}
\end{center}
\caption{(a) A schematic picture of one-dimensional chain which is created
by adding sites from left to right. (b) A schematic picture of
one-dimensional chain which is created by stacking sites both from left to
right and right to left and finally connect the two chains. }
\label{fig1}
\end{figure}

First, we review some general aspects of the recursive Green's function
technique. Its spirit is as follows: to build the full Green's function, we
start from isolated blocks and connect them to other parts of the system by
stacking sites one by one along a certain direction. For example, we
consider a one-dimensional atomic chain along $x$-direction as shown in
Fig. \ref{fig1}(a) and suppose that the Green's function of the detached $N$ sites on the
left side has been known. We denote $G_{N,N}^{\text{L}}$ as the Green's
function at the $N$ site. Consider adding one more site from the right to this
system. Based on Dyson equation, we have
\begin{equation}
G_{N+1,N+1}^{\text{L}}=g_{\text{iso}}+g_{\text{iso}}V_{N+1,N}G_{N,N+1}^{%
\text{L}}  \label{eq1}
\end{equation}%
\begin{equation}
G_{N,N+1}^{\text{L}}=G_{N,N}^{\text{L}}V_{N,N+1}G_{N+1,N+1}^{\text{L}},
\label{eq2}
\end{equation}%
where $g_{\text{iso}}$ is the Green's function of the isolated $N+1$ site.
Substituting Eq. (\ref{eq1}) into Eq. (\ref{eq2}), we get
\begin{equation}
G_{N+1,N+1}^{\text{L}}=\Bigl[g_{\text{iso}}^{-1}-V_{N+1,N}G_{N,N}^{\text{L}%
}V_{N,N+1}\Bigr]^{-1}.  \label{eq3}
\end{equation}%
LDOS can be calculated as follows:
\begin{equation}
\rho ^{L}(E)=-\frac{1}{\pi }\text{ImTr}\Bigl[G_{N+1,N+1}^{\text{L}}(E+i0^{+})%
\Bigr],  \label{LDOS}
\end{equation}%
with infinitesimal positive number $0^{+}$. Similarly, one can determine the
Green's function by stacking from right to left (in this case the
superscript of $G$ is replaced with $R$, such as $G_{N,N}^{\text{R}}$).
Provided that we start this stacking process simultaneously from the two ends,
we can obtain the Green's function $G_{N-1,N-1}^{\text{L}}$ at site $N-1$ of
the left chain and $G_{N+2,N+2}^{\text{R}}$ at site $N+2$ of the right
chain.
From Eq. (\ref{eq3}), we know that the process of adding the site $N$ in the
left chain generates
\begin{equation}
G_{N,N}^{\text{L}}=\Bigl[g_{\text{iso}}^{-1}-V_{N,N-1}G_{N-1,N-1}^{\text{L}%
}V_{N-1,N}\Bigr]^{-1},  \label{eq4}
\end{equation}%
and similarly
\begin{equation}
G_{N+1,N+1}^{\text{R}}=\Bigl[g_{\text{iso}}^{-1}-V_{N+1,N+2}G_{N+2,N+2}^{%
\text{R}}V_{N+2,N+1}\Bigr]^{-1},  \label{eq5}
\end{equation}%
from adding the site $N+1$ in the right chain. Now we connect the two chains
(light-blue dashed line in Fig. \ref{fig1}(b)) and denote the Green's function of
this combined chain as $G$. Based on Dyson equation we get following
equations:
\begin{eqnarray}
&&G_{N,N}=\Bigl[G_{N,N}^{-1\text{L}}-V_{N,N+1}G_{N+1,N+1}^{\text{R}}V_{N+1,N}%
\Bigr]^{-1},  \label{eq6} \\
&&G_{N+1,N+1}=\Bigl[G_{N+1,N+1}^{-1\text{R}}-V_{N+1,N}G_{N,N}^{\text{L}%
}V_{N,N+1}\Bigr]^{-1}  \label{eq7} \\
&&G_{N,N+1}=G_{N,N}^{\text{L}}V_{N,N+1}G_{N+1,N+1} \label{eq8} \\
&&G_{N+1,N}=G_{N+1,N+1}^{\text{R}}V_{N+1,N}G_{N,N}.  \label{eq9}
\end{eqnarray}%
Suppose hopping amplitude of adjacent sites is given by $t$, we can calculate the current
\begin{equation}
J=-ietk_BT\sum_{\omega _{n}}\text{Tr}\Bigl[G_{N+1,N}(\omega
_{n})-G_{N,N+1}(\omega _{n})\Bigr],  \label{current}
\end{equation}%
with Boltzman constant $k_B$, temperature $T$, and Matsubara frequency $\omega _{n}$.

\begin{figure}[h]
\begin{center}
\includegraphics[width=80mm]{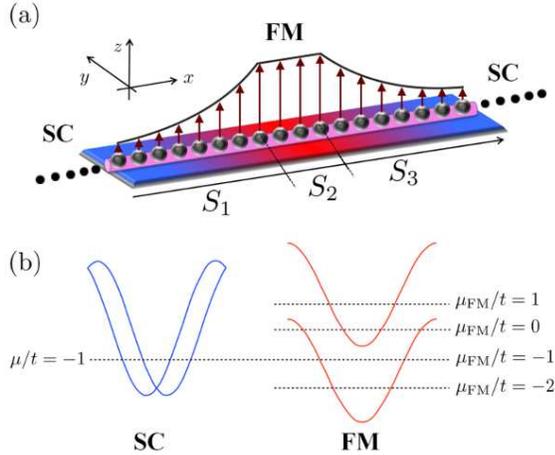}
\end{center}
\caption{(a) A schematic picture of a nanowire on top of S/F/S junction. (b)
Chemical potential of a nanowire in the superconductor segment (left) and
ferromagnet segment (right). }
\label{fig2}
\end{figure}

Next, we construct the model Hamiltonian of the semiconducting nanowire on
top of the S/F/S junction (Fig. \ref{fig2}(a)). We separate this nanowire into three
parts by introducing
\begin{eqnarray}
S_{1} &=&\bigl\{n|1\leq n\leq L_{\text{SC}}\bigr\}  \nonumber \\
S_{2} &=&\{n|L_{\text{SC}}+1\leq n\leq L_{\text{SC}}+L_{\text{FM}}\}
\nonumber \\
S_{3} &=&\{n|L_{\text{SC}}+L_{\text{FM}}+1\leq n\leq 2L_{\text{SC}}+L_{\text{%
FM}}\}
\end{eqnarray}%
with site index $n$. We denote the equal site length of each side of
superconductors as $L_{\text{SC}}$ and that of ferromagnet as $L_{\text{FM}}$%
. We define Hamiltonian as follows:
\begin{widetext}
\begin{eqnarray}
&&\mathcal{H}=-t\sum_{\substack{\left\langle m,n\right\rangle\\ \sigma}}c_{m\sigma}^{\dagger}c_{n\sigma}+\sum_{\substack{n\in S_1\\n\in S_3}}(\frac{A}{2}c_{n\uparrow}^{\dagger}c_{n+1\downarrow}-\frac{A}{2}c_{n\downarrow}^{\dagger}c_{n+1\uparrow}+\text{H.c.})
-\sum_{\substack{n\in S_1\\ \sigma}}\mu c_{n\sigma}^{\dagger}c_{n\sigma}+\sum_{n\in S_1}(\Delta c_{n\uparrow}^{\dagger}c_{n\downarrow}^{\dagger}+\text{H.c.})\notag\\
&&-\sum_{\substack{n\in S_3\\ \sigma}}\mu c_{n\sigma}^{\dagger}c_{n\sigma}+\sum_{n\in S_3}(\Delta e^{i\varphi} c_{n\uparrow}^{\dagger}c_{n\downarrow}^{\dagger}+\text{H.c.})
-\sum_{\substack{n\in S_2\\ \sigma}}\mu_{\text{FM}}c_{n\sigma}^{\dagger}c_{n\sigma}+\sum_{n\in S_2}V_z(c_{n\uparrow}^{\dagger}c_{n\uparrow}-c_{n\downarrow}^{\dagger}c_{n\downarrow})\notag\\
&&+\sum_{n\in S_1}V_z\exp\Bigl[\frac{n-L_{\text{SC}}}{\xi_L}\Bigr](c_{n\uparrow}^{\dagger}c_{n\uparrow}-c_{n\downarrow}^{\dagger}c_{n\downarrow})
+\sum_{n\in S_3}V_z\exp\Bigl[\frac{L_{\text{SC}}+L_{\text{FM}}+1-n}{\xi_R}\Bigr]
(c_{n\uparrow}^{\dagger}c_{n\uparrow}-c_{n\downarrow}^{\dagger}c_{n\downarrow})\label{model}
\end{eqnarray}
\end{widetext}where $c_{n\sigma }^{\dagger }(c_{n\sigma })$ is the electron
creation (annihilation) operator with site $n$ and spin $\sigma $, $t$ is
the hopping matrix between  the nearest neighbor $\left\langle
i,j\right\rangle $, $A$ is the Rashba spin orbit coupling (SOC), $\mu (\mu _{%
\text{FM}})$ is chemical potential in superconductor (ferromagnet) segment, $%
\Delta $ is the pair potential, $\varphi $ is the phase difference of
superconductors, $V_{z}$ is the ferromagnetic order, and $\xi _{R(L)}$ is
the decay length of ferromagnetic order in the right (left) superconductor.
Unlike the model Hamiltonian on top of the superconductor with uniform
ferromagnetic order, in this model construction, we assume that
ferromagnetic order penetrates into right (left) superconductor segment
decaying as $\sim \exp [\frac{-n}{\xi _{R}}](\exp [\frac{n}{\xi _{L}}])$ as
described in Eq. (\ref{model}). As for Rashba SOC, we do not include the
second term in Eq. (\ref{model}) at the interface between superconductor and
ferromagnet since it does not affect our results.

Now we apply the recursive Green's function technique to calculate LDOS,
Josephson current and ABS. The retarded and Matsubara Green's function of the
isolated site can be described as
\begin{equation}
g_{\text{iso}}(E)=\frac{1}{(E+i0^{+})-\mathcal{H}_{\text{iso}}}  \label{eq8}
\end{equation}%
\begin{equation}
g_{\text{iso}}(\omega _{n})=\frac{1}{i\omega _{n}-\mathcal{H}_{\text{iso}}}
\label{eq9}
\end{equation}%
where $\mathcal{H}_{\text{iso}}$ is
\[
\mathcal{H}_{\text{iso}}=c_{n}^{\dagger }\hat{\mathcal{H}}_{\text{iso}}c_{n}
\]%
\begin{equation}
\hat{\mathcal{H}}_{\text{iso}}=\begin{cases}
-\mu\sigma_0\tau_z-\sigma_y\tau_y\Delta, \;\;n\in S_1\\
-\mu_{\text{FM}}\sigma_0\tau_z, \;\;n\in S_2\\
-\mu\sigma_0\tau_z-\sigma_y\tau_{\downarrow}\Delta
e^{i\varphi}-\sigma_y\tau_{\uparrow}\Delta e^{-i\varphi}, \;\;n\in S_3
\end{cases}
\end{equation}%
with the basis $c_{n}=(c_{n\uparrow },c_{n\downarrow },c_{n\uparrow
}^{\dagger },c_{n\downarrow }^{\dagger })$. $\sigma _{0,x,y,z}(\tau
_{0,x,y,z})$ is Pauli matrix in spin (particle-hole) space and $\tau
_{\uparrow }=(\tau _{x}+i\tau _{y})/2$, $\tau _{\downarrow }=(\tau
_{x}-i\tau _{y})/2$. Hopping matrix can be written as follows:
\begin{eqnarray*}
V_{n,n+1} &=&c_{n}^{\dagger }\hat{V}_{n,n+1}c_{n+1} \\
V_{n+1,n} &=&c_{n+1}^{\dagger }\hat{V}_{n+1,n}c_{n},
\end{eqnarray*}%
\begin{equation}
\hat{V}_{n,n+1}=\begin{cases} -t\sigma_0\tau_z-i\frac{A}{2}\tau_z\sigma_y,
\;\;n\in S_1, n\in S_3\\ -t\sigma_0\tau_z, \;\;n\in S_2 \end{cases}
\end{equation}%
\begin{equation}
\hat{V}_{n+1,n}=\begin{cases} -t\sigma_0\tau_z+i\frac{A}{2}\tau_z\sigma_y,
\;\;n\in S_1, n\in S_3\\ -t\sigma_0\tau_z, \;\;n\in S_2 \end{cases}
\end{equation}%
In the next section, we will calculate the LDOS on the edge of the nanowire
with proximity coupled
pair potential and decaying ferromagnetic order, Josephson
current and ABS of the nanowire on S/F/S junction.
LDOS is given by
\begin{equation}
\rho ^{L}(E)=-\frac{1}{\pi }\text{ImTr}\Bigl[G_{L_{\text{SC}},L_{\text{SC}%
}}^{L}(E+i0^{+})\Bigr].  \label{LDOS}
\end{equation}%
From Eq. (\ref{current}), Josephson current and ABS are described as follows:
\begin{equation}
J(\varphi )=-ietk_BT\sum_{\omega _{n}}\text{Tr}\Bigl[G_{n+1,n}(\omega
_{n})-G_{n,n+1}(\omega _{n})\Bigr],\;\;n\in S_{2}  \label{Josephson}
\end{equation}%
\begin{equation}
\rho (E,\varphi )=-\frac{1}{\pi }\text{ImTr}\Bigl[G_{n,n}(E+i0^{+})\Bigr]%
,\;\;n\in S_{2}  \label{ABS}
\end{equation}%
where $\varphi$ is the macroscopic phase difference of pair potential
between two superconductors.
We can calculate $\varphi$ dependence explicitly.

\section{Results}
This section consists of four parts: in subsection A, we study the LDOS on the edge of
the nanowire with both pair potential and decaying ferromagnetic
order and exhibit the evolution of the surface resonance modes. In subsection B, we then
calculate Josephson current S/F/S junctions of the nanowire with several
different decay length. It shows that at specific parameter tuning,
anomalous current-phase relation which can be roughly regarded as $\sin
2\varphi $ appears. In subsection C, we calculate ABS.
From the spectrum of ABS, we will provide a simple
argument of explanation that the emergence of crossing points of ABS is
the important factor to realize the current-phase relation $\sin 2\varphi $.
In subsection D, we finally analyze the symmetry of pair amplitudes in this junction especially focusing on odd-frequency pairing. Odd-frequency spin-triplet
pairing is enhanced when the Majorana like zero energy state is generated at
the S/F (F/S) interface.\cite{Asano2013} Throughout this section, we fix the parameters as follows: $%
\Delta /t=0.1,\mu /t=-1,A/t=1,V_{z}/t=1.5$. With this choice of parameters, condition of topological non-trivial state\cite{oreg10} is satisfied when ferromagnetic order is {\it uniform}. We set the number of sites of superconductor segment long enough so that the effect of overlapping between two zero energy modes on the edges of segment is negligible. In actual numerical calculation,
we fix the number of sites
of superconductor segment as $L_{\text{SC}}=4000$ and ferromagnet one as
$L_{\text{FM}}=4$. To calculate retarded Green's function, we set the infinitesimal positive number $0^+$ as $0^+/t=0.001$.

\subsection{LDOS}\label{sec3a}
\begin{figure}[h]
\begin{center}
\includegraphics[width=80mm]{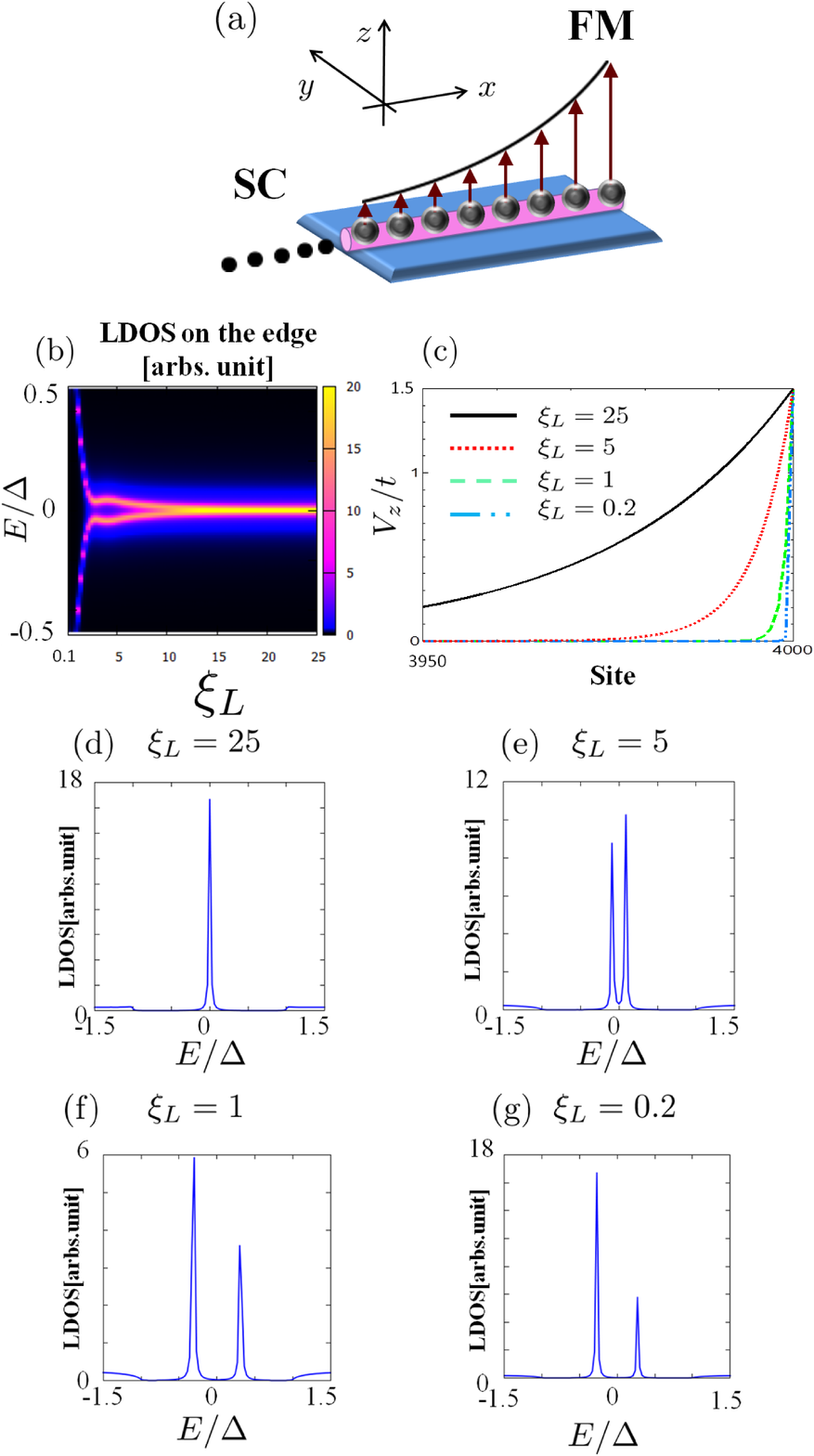}
\end{center}
\caption{(a) A schematic picture of a nanowire on top of superconductor with
decaying ferromagnetic order. (b) The intensity plot of LDOS on the edge ($%
i.e.$ $n=L_{\text{SC}}$) of the nanowire shown in (a). (c) The spatial
profiles of ferromagnetic order with different decay length, $\protect\xi %
_{L}=25$ (black solid line), $\protect\xi _{L}=5$ (red dotted line), $%
\protect\xi _{L}=1$ (green dashed line), and $\protect\xi _{L}=0.2$ (blue 2
dotted line). (d)-(g) LDOS on the edge of the nanowire with different
decay length. }
\label{fig3}
\end{figure}

In this subsection, we examine the LDOS on the edge of nanowire which is
calculated from Eq. (\ref{LDOS}). The situation is illustrated in Fig. \ref{fig3}(a),
where the pair potential and decaying ferromagnetic order coexist.
The intensity of LDOS is plotted in Fig.  \ref{fig3}(b). It is shown that for the
long decaying of $V_{z}$  $\left( \xi _{L}>10\right) $, there is a single
resonance peak at zero energy. This is in
agreement with the finding that in the asymptotic scenario when $V_{z}$ is spatially
\textit{uniform} at infinite $\xi _{L}$, such nanowire is topologically
non-trivial which can be confirmed by the sign of Pfaffian with our choice
of parameters.\cite{Kitaev01}
However, in the presence of spatially \textit{nonuniform} ferromagnetic order, the
preexisting topological argument is no longer valid. Thus, it is remarkable
to see here that even if the ferromagnetic order is non-uniform, LDOS with
zero energy peak can still appear on the edge of the nanowire within the
numerical accuracy. On the opposite, with the decay length less than around $%
\xi _{L}\sim 10$,  this zero mode splits into two resonance peaks symmetric
to zero energy. Owing to the localized ferromagnet at the end of nanowire, this
finding associates with another important physical results known as Shiba
states.\cite{Yu,Shiba,Rusinov} It is well known that when a magnetic
impurity is put on the superconductor, there is a bound state around the
impurity inside the superconducting energy gap. In the present model, we can
control the ``zero mode-non-zero mode crossover" with decreasing the decay
length of ferromagnetic order. In Figs. \ref{fig3}(d)-(g), we plot the LDOS for
a selected decay length $\xi _{L}=25,5,1,$ and $0.2$ respectively (see also
Fig. \ref{fig3}(c) where the spatial profiles of ferromagnetic order are shown for
four different decay length). When $\xi _{L}=25$, we see the zero energy
mode, on the other hand, in the rest of the cases (Figs. \ref{fig3}(e)-(g)), we find non-zero modes in
the energy gap of superconducting region.\par
In Appendix \ref{ap1}, we consider the overlapping of two zero energy modes in the shorter length system and also compare non-uniform ferromagnetic case and uniform case, which leads that two zero energy modes appeared in non-uniform situation can be regarded as two MFs.
\subsection{Josephson current}\label{sec3b}

In this subsection, we study Josephson currents for various decay lengths of
ferromagnetic order on both sides of the superconductor. On the left side,
we are interested in three typical decay lengths $\xi _{L}=25,1,$ and $0$ ($%
i.e.$, ferromagnetic order does not penetrate into the left superconductor).
In each case, we will tune the decay length $\xi _{R}$ as well as the
chemical potential of the ferromagnet segment ($\mu _{\text{FM}}/t=-2,-1,0,$
and $1$, see also Fig. \ref{fig2}(b)).
\begin{figure}[h]
\begin{center}
\includegraphics[width=60mm]{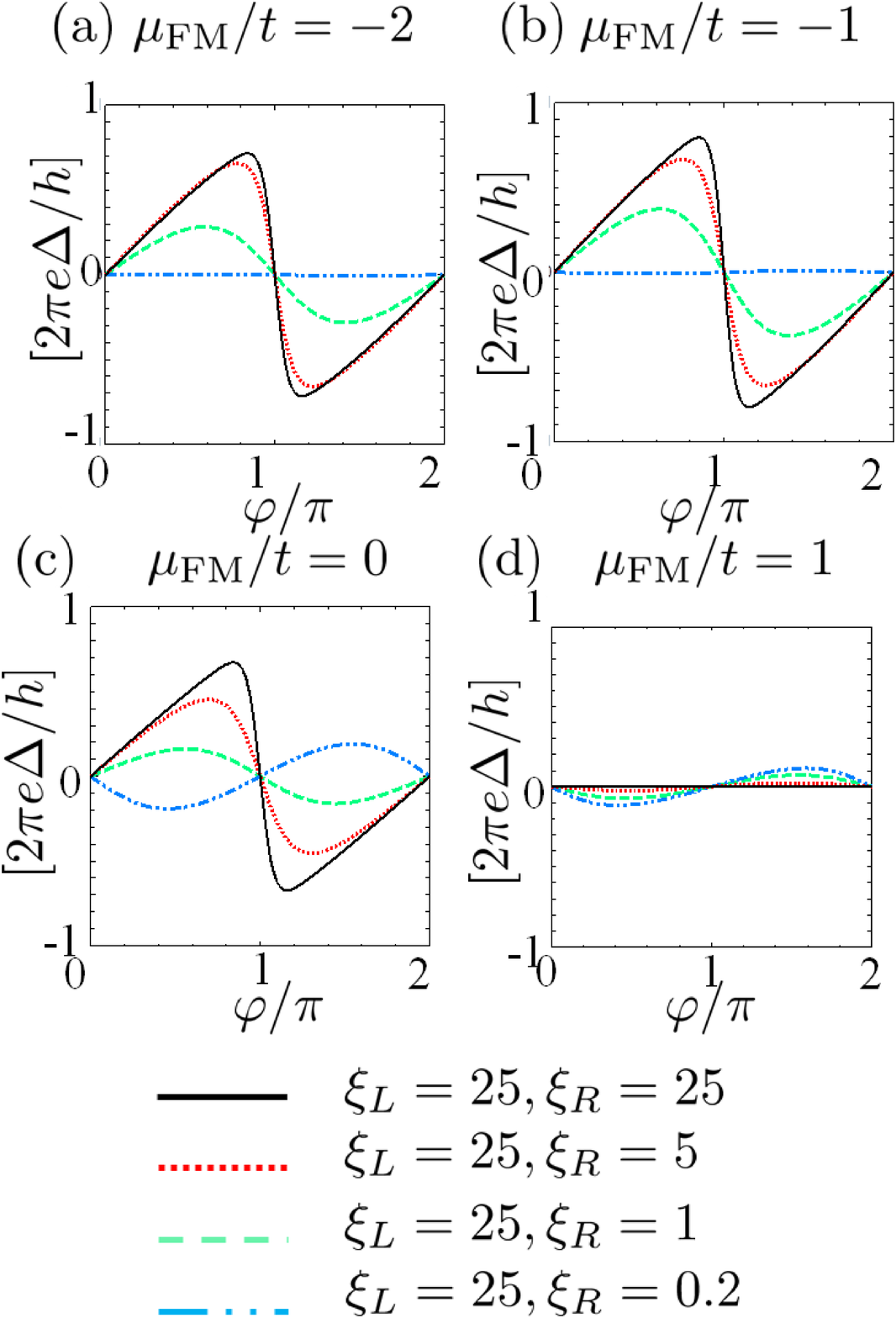}
\end{center}
\caption{(a)-(d) Josephson current of the nanowire on top of S/F/S
junction in the case of $\protect\xi _{L}=25$ with four different chemical
potentials of ferromagnetic layer listed above each figure. There are four
lines in each figure: $\protect\xi _{L}=25,\protect\xi _{R}=25$ (black solid
line), $\protect\xi _{L}=25,\protect\xi _{R}=5$ (red dashed line), $\protect%
\xi _{L}=25,\protect\xi _{R}=1$ (green dashed line), and $\protect\xi %
_{L}=25,\protect\xi _{R}=25$ (blue 2 dotted line).}
\label{fig4}
\end{figure}

First, we look at the case $\xi _{L}=25$. As the previous discussion shows,
there is zero energy mode on the edge of the left nanowire. In Figs. \ref{fig4}(a)-(d), we plot Josephson currents for $\mu _{\text{FM}}/t=-2,-1,0,$ and $%
1$, respectively. In each figure, we take four values of $\xi _{R}$: $25$, $5
$, $1$ and $0.2$. The phase dependence of the current in all cases has the
dominant coupling proportional to $\sin \varphi $. Interestingly, for $\mu _{%
\text{FM}}/t=-2,-1$ and $0$, the Josephson current in symmetric junctions
abruptly changes its sign at $\varphi =\pi $ as shown in Figs. \ref{fig4}(a), (b) and
(c), respectively. In asymmetric junctions, such jump is absent. Notice that
our system considered here has no perfect transmissivity and is closely
similar to that of two $d$-wave superconductor with zero energy ABSs,\cite{Kashiwaya97,TK96j,Kashiwaya00}
$p$-wave superconductors \cite{Yakovenko,Asano2006}
or Kitaev chains junction
system where MFs are coupled each other.\cite{Kitaev01,JosephsonMF}
Therefore, the abrupt jump can only be
explained by the existence of robust zero energy
ABSs,\cite{ABS,ABSb,Hu,TK95,TK96a}
$i.e.$, MFs.\cite{Kitaev01}
However, in the non-uniform ferromagnetic order, the formation of zero
modes are distinct from $p$-wave superconductor or Kitaev model. It is
interesting that the similar behavior of Josephson current is found even in
the present model. For $\mu _{\text{FM}}/t=1$, we obtain $\pi $-state and
the current are quite small compared to the other cases of chemical
potentials. In this case, there is one band at the chemical potential in ferromagnetic
region which has opposite spin compared to the ferromagnetic order in SCs.
Thus, the transparency of the junction is greatly reduced and $\pi $-state
can appear as a result of misaligned magnetizations.

\begin{figure}[h]
\begin{center}
\includegraphics[width=60mm]{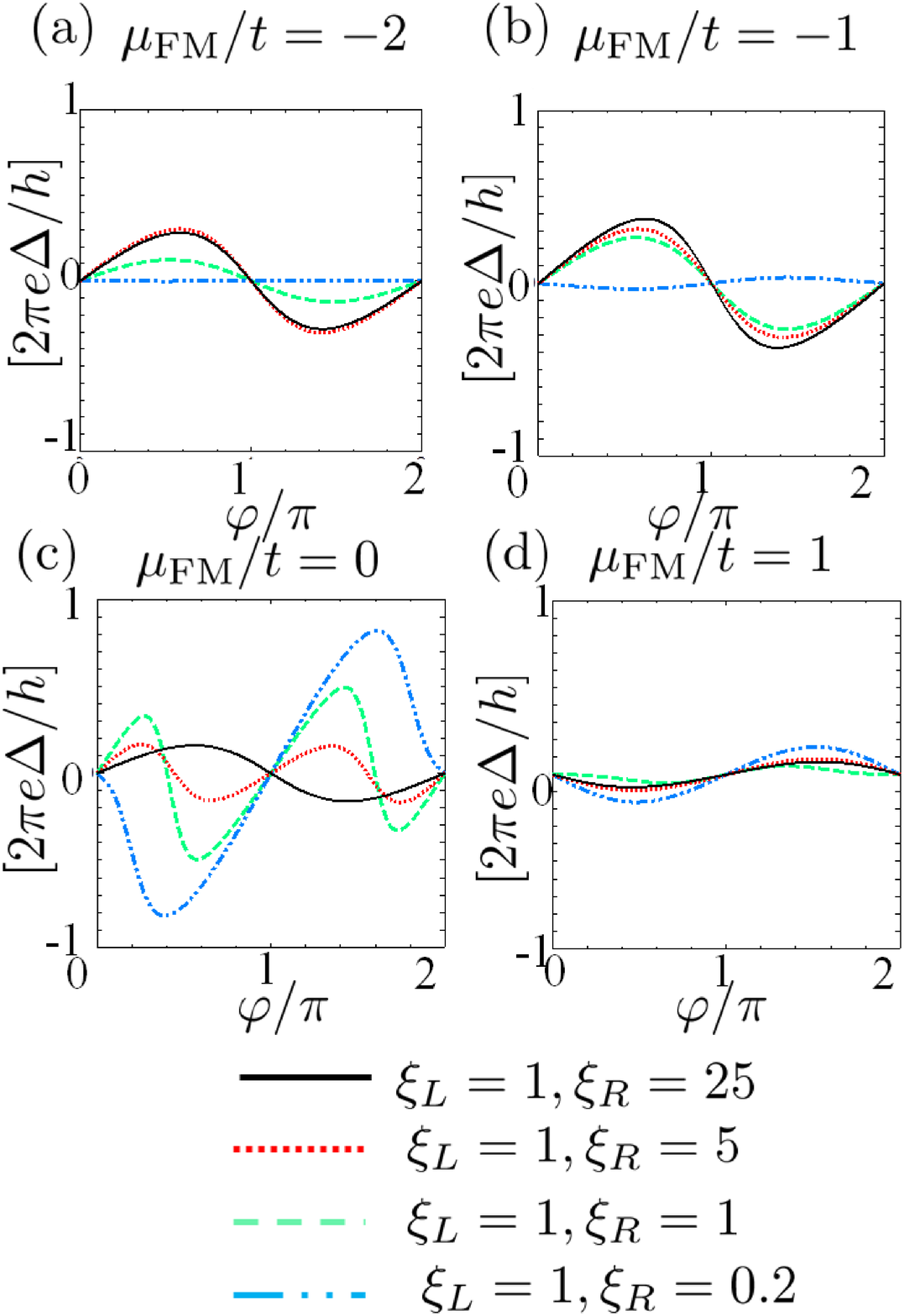}
\end{center}
\caption{(a)-(d) Josephson current of the nanowire on top of S/F/S
junction in the case of $\protect\xi _{L}=1$ with four different chemical
potentials of ferromagnetic layer listed in above each figure. There are
four lines in each figure: $\protect\xi _{L}=1,\protect\xi _{R}=25$ (black
solid line), $\protect\xi _{L}=1,\protect\xi _{R}=5$ (red dashed line), $%
\protect\xi _{L}=1,\protect\xi _{R}=1$ (green dashed line), and $\protect\xi %
_{L}=1,\protect\xi _{R}=0.2$ (blue 2 dotted line). }
\label{fig5}
\end{figure}

Second, we focus on the case $\xi _{L}=1$ where non-zero modes are localized
on the left segment. As for $\mu _{\text{FM}}/t=-2$ and $-1$ (Figs. \ref{fig5}(a) and
(b)), we find the current-phase relation $\sin \varphi $ in three cases: $%
\xi _{R}=25$ (black solid line), $\xi _{R}=5$ (red dotted line), and $\xi _{R}=1$ (green dashed line).
On the other hand Josephson current is almost zero when $\xi _{R}=0.2$
(blue 2 dashed line). Surprisingly, when $\mu _{\text{FM}}/t=-2$, $\xi _{R}=5$ and $%
\xi _{R}=1$ (Fig. \ref{fig5}(c) red dotted and green dashed lines), we find the anomalous
current-phase relation which can be roughly regarded as $\sin 2\varphi $. As
we will see in the next subsections, the coupling of non-zero modes produces
this anomalous current-phase relation especially when the emergence of
crossing points of ABS becomes the important factor. For $\mu _{\text{FM}%
}/t=1$, all of the Josephson currents are suppressed compared to the other
cases of chemical potential.

\begin{figure}[h]
\begin{center}
\includegraphics[width=60mm]{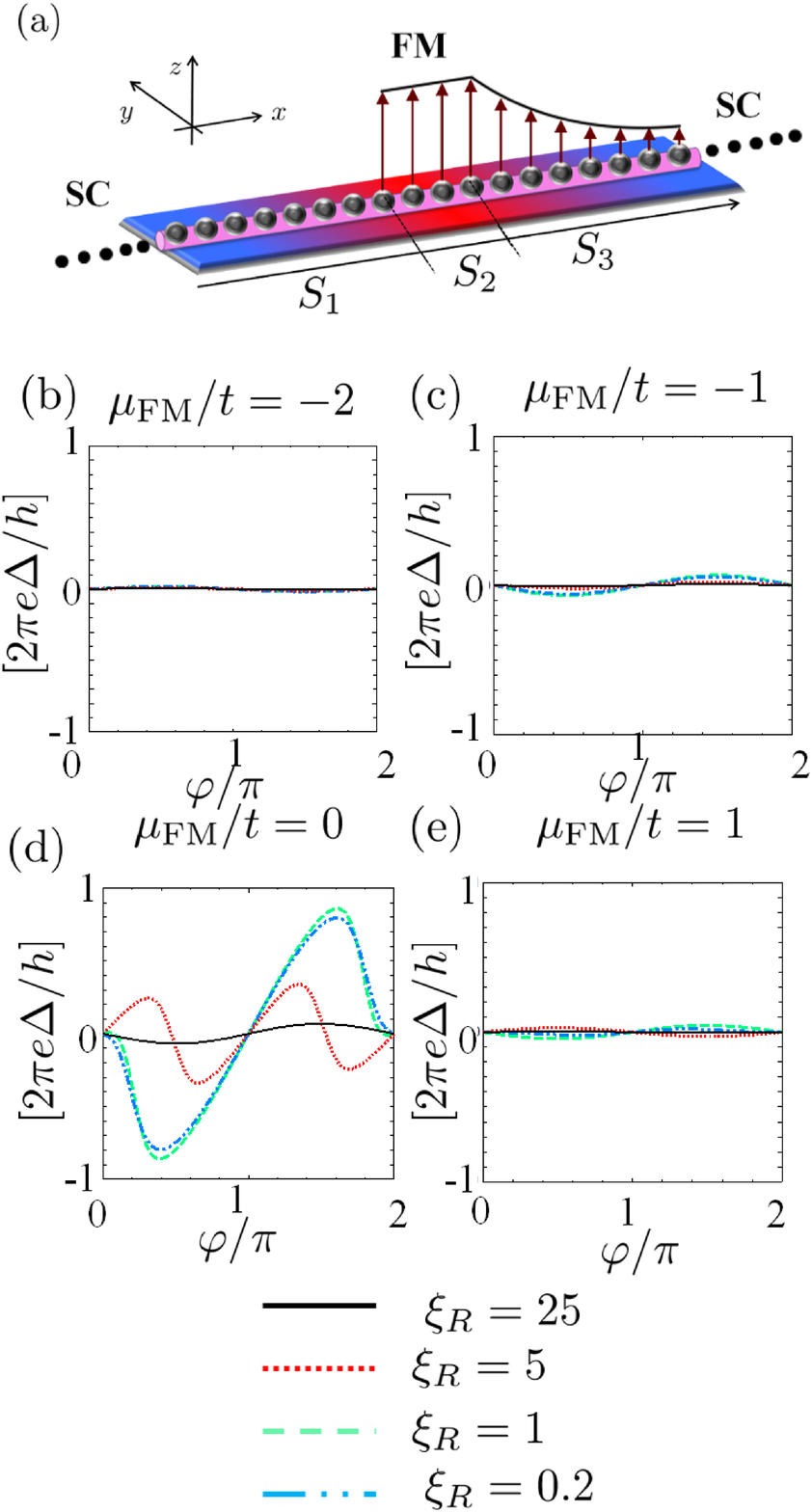}
\end{center}
\caption{(a) A schematic picture of a nanowire on top of S/F/S junction. (a)-(d) Josephson current of the nanowire on top of S/F/S junction in the
case of $\protect\xi _{L}=0$, that is, ferromagnetic order does not
penetrate into left superconductor segment. Results are shown with four
different chemical potentials of ferromagnetic layer listed above each
figure. There are four lines in each figure: $\protect\xi _{L}=0,\protect\xi %
_{R}=25$ (black solid line), $\protect\xi _{L}=0,\protect\xi _{R}=5$ (red
dashed line), $\protect\xi _{L}=0,\protect\xi _{R}=1$ (green dashed line),
and $\protect\xi _{L}=0,\protect\xi _{R}=25$ (blue 2 dotted line). }
\label{fig6}
\end{figure}

Finally, we study the case $\xi _{L}=0$, $i.e.$, the ferromagnetic order
does not penetrate into the left superconductor segment. When $\mu _{\text{FM%
}}/t=-2,-1$ and $1$ (Figs. \ref{fig6}(b), (c), and (d), respectively), Josephson
current is almost zero, on the other hand, when $\mu _{\text{FM}}/t=0$, we
see current-phase relation $\sin 2\varphi $ with decay length on the right $%
\xi _{R}=5$ (dotted red line) and $-\sin \varphi $ with other decay length.

Before we proceed, it is instructive to summarize the interesting phenomena
we have found in this subsection. In the case of $\xi _{L}=25$ and $\mu _{%
\text{FM}}/t=-2,-1,0$, we see the abrupt sign reversal of current at $%
\varphi =\pi $. In the case of $\xi _{L}=25$ and $\mu _{\text{FM}}/t=1$, the
amplitude of Josephson current is suppressed. When $\xi _{L}=1$, $\mu _{%
\text{FM}}/t=0$, and $\xi _{R}=5$ or $1$, we obtain current-phase relation
approximated as $\sin 2\varphi $. We also find this current-phase relation $%
\sin 2\varphi $ in the case of $\xi _{L}=0$, $\mu _{\text{FM}}/t=0$, and $%
\xi _{R}=5$.


\subsection{ABS}\label{sec3c}

In this subsection, we study the ABS of nanowire on S/F/S junction with
different decay length and relate it to the behavior of Josephson current
obtained in the previous subsection. We mainly focus on the case of $\mu _{%
\text{FM}}/t=0$. It is well known that when the magnitudes of the pair
potential are the same on the left and right side of superconductor,
Josephson current can be calculated by:
\begin{equation}
J(\varphi )=\frac{2e}{h}\frac{\partial F(\varphi )}{\partial \varphi },
\end{equation}%
where $F$ is free energy. In one dimension, $F$ can be written as
\begin{equation}
F=-k_{B}T\sum_{n}\log \Bigl(2\cosh \bigl(\frac{\varepsilon _{n}}{2k_{B}T}%
\bigr)\Bigr)
\end{equation}%
where $\varepsilon _{n}$ is the energy of
ABS \cite{Beenakker13}. Therefore, Josephson current is
\begin{eqnarray}
&&J(\varphi )=-\frac{2e}{h}k_{B}T\sum_{n}\frac{1}{\cosh \bigl(\frac{%
\varepsilon _{n}}{2k_{B}T}\bigr)}\sinh \bigl(\frac{\varepsilon _{n}}{2k_{B}T}%
\bigr)\frac{1}{2k_{B}T}\frac{\partial \varepsilon _{n}}{\partial \varphi }
\nonumber \\
&&=-\sum_{n}\frac{e}{h}\tanh \bigl(\frac{\varepsilon _{n}}{2k_{B}T}\bigr)%
\frac{\partial \varepsilon_n }{\partial \varphi }  \label{current}
\end{eqnarray}%
At low temperature, $J(\varphi )$ can be approximated as
\begin{equation}
J(\varphi )\sim -\frac{e}{h}\sum_{n}\text{sgn}(\varepsilon _{n})\frac{%
\partial \varepsilon _{n}}{\partial \varphi }.  \label{apprx}
\end{equation}%
sgn$(\varepsilon _{n})$ gives $+1(-1)$ when $\varepsilon _{n}$ is positive
(negative). In the above, $n$ denotes the band index of ABS. Due to the
particle-hole symmetry, we can only take the ABSs below the zero energy into
account. Thus, Josephson current can be approximated as the derivative of
ABSs below the zero energy.

\begin{figure}[h]
\begin{center}
\includegraphics[width=80mm]{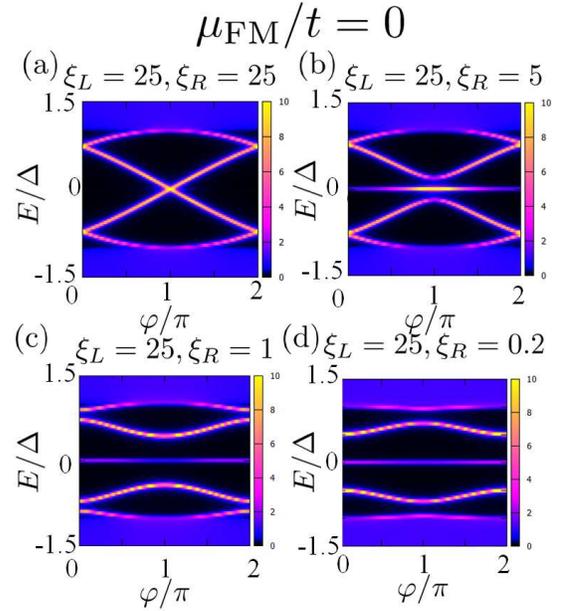}
\end{center}
\caption{(a)-(d) ABS (the intensity plot of LDOS at the ferromagnet segment within the
superconducting energy gap) in the case of $\protect\mu_{\text{FM}}/t=0$ and
$\protect\xi_L=25$ with different decay length $\protect\xi_R$. }
\label{fig7}
\end{figure}

\begin{figure}[h]
\begin{center}
\includegraphics[width=80mm]{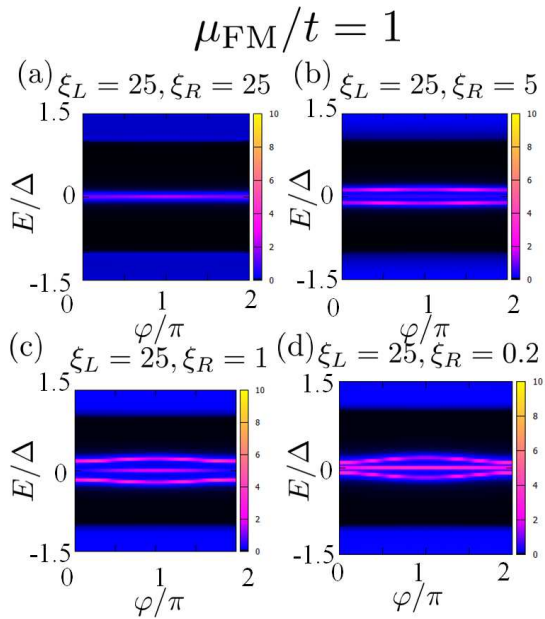}
\end{center}
\caption{(a)-(d) ABS (the intensity plot of LDOS at the ferromagnet segment within the
superconducting energy gap) in the case of $\protect\mu _{\text{FM}}/t=1$
and $\protect\xi _{L}=25$ with different decay length $\protect\xi _{R}$. }
\label{fig8}
\end{figure}
First, we look at the case $\xi _{L}=\xi _{R}=25$ where the zero
energy modes are located on both sides. These two zero modes hybridize as
indicated in Fig. \ref{fig7}(a). The crossing point at $\varphi =\pi $ explains the
sudden drop of Josephson current which can lead to the unusual $4\pi $
periodicity of current-phase relation if we consider AC Josephson current.
As the decay length on the right decreases, the zero energy mode on the left
does not hybridize with the states on the right, which can be seen from the
flat ABS as a function of $\varphi $ in Figs. \ref{fig7}(b)-(d). The
contribution of Josephson current is mainly carried by the ABSs away from zero
energy. With this estimation, we can relate ABS to the behavior of Josephson
current in the previous subsection. If we look at  Figs. \ref{fig7}(b) and (c)
((d)), ABSs below the zero energy change as $\sim -\cos \varphi $ ($\sim
\cos \varphi $) which leads to current-phase relation $\sin \varphi $ ($-\sin
\varphi $). This consideration corresponds well to the black solid line and
red dotted line (green dashed line) in Fig. \ref{fig4}(c). If we tune the chemical
potential of ferromagnetic layer as $\mu _{\text{FM}}/t=1$, the ABSs are
almost flat which do not make contribution to the Josephson current (Fig. \ref{fig8}).
Indeed, the amplitude of Josephson currents are almost zero as shown in
Fig. \ref{fig4}(d).

\begin{figure}[h]
\begin{center}
\includegraphics[width=80mm]{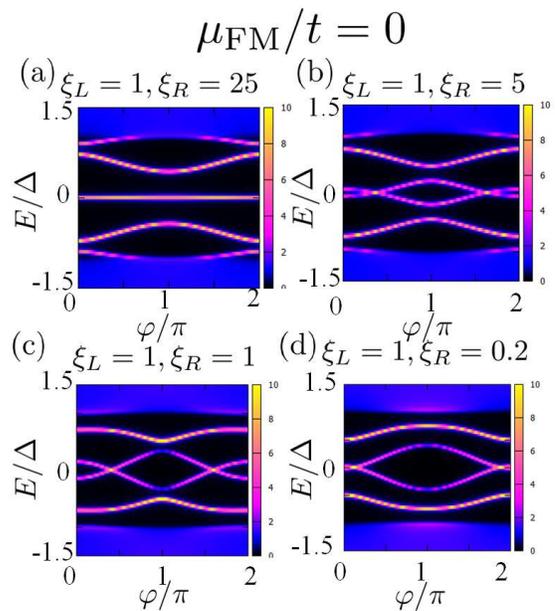}
\end{center}
\caption{(a)-(d) ABS (the intensity plot of LDOS at the ferromagnet segment within the
superconducting energy gap) in the case of $\protect\mu _{\text{FM}}/t=0$
and $\protect\xi _{L}=1$ with different decay length $\protect\xi _{R}$. }
\label{fig9}
\end{figure}
Next, we focus on the case $\xi _{L}=1$, when non-zero modes are localized at
the interface between left superconductor and ferromagnet segment. When
decay length $\xi _{R}=25$ (Fig. \ref{fig9}(a)), the zero energy mode is located on the
right segment. This mode is not coupled with the left non-zero modes. The
major change of ABSs below the zero energy can be regarded as $\sim
-\cos \varphi $ (Fig. \ref{fig9}(a)) which contributes $\sin \varphi $ to the Josephson
current (Fig. \ref{fig5}(c) black solid line).

In the case of $\xi _{R}=5$ and $\xi _{R}=1$, when the anomalous current-phase relation $\sin 2\varphi $ can be seen, we find that non-zero modes on
the both sides hybridize and these states are crossed
at two values of $\varphi$:
one is located at between $\varphi =0$
and $\pi $, another is between $\varphi =\pi $ and $\varphi =2\pi $ (Figs. \ref{fig9}(b) and (c)).
As we discuss later, this crossing points of ABS is important to realize
current-phase relation $\sin 2\varphi $. For $\xi _{R}=0.2$, all of the ABSs
change as $\cos \varphi $ (Fig. \ref{fig9}(d)) which gives current-phase relation $%
\sin \varphi $ (Fig. \ref{fig5}(c) blue 2 dotted line).

\begin{figure}[h]
\begin{center}
\includegraphics[width=80mm]{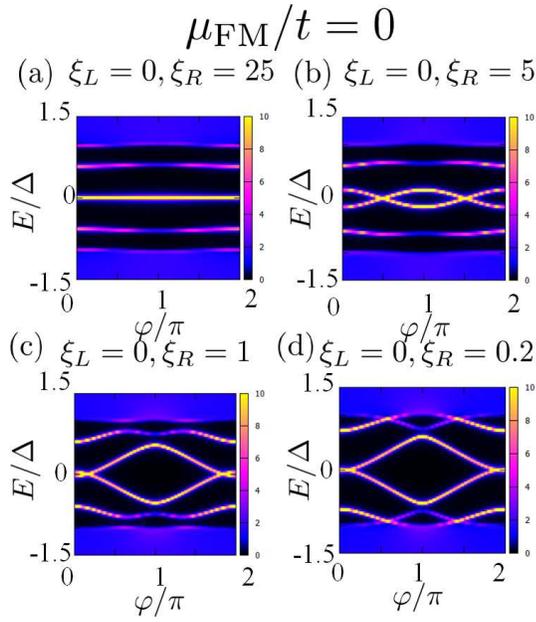}
\end{center}
\caption{(a)-(d) ABS (the intensity plot of LDOS at the ferromagnet segment within the
superconducting energy gap) in the case of $\protect\mu _{\text{FM}}/t=0$
and $\protect\xi _{L}=0$ ($i.e.$ there is no ferromagnetic order in the left
superconductor segment.) with different decay length $\protect\xi _{R}$. }
\label{fig10}
\end{figure}
Finally, we focus on the case $\xi _{L}=0$, $i.e.$, ferromagnetic order is
absent in the left superconductor side (Fig. \ref{fig10}). If we set $\xi _{R}=25$,
all of the ABSs becomes flat as a function of $\varphi $ which leads to
almost zero Josephson current (Fig. \ref{fig6}(c) black solid line). For $\xi
_{R}=1,0.2$, the change of ABS below the zero energy obeys $\sim \cos
\varphi $ (Figs. \ref{fig10}(c) and (d)), thus, current-phase relation is $-\sin
\varphi $ (Fig. \ref{fig6}(c) green dashed line and blue 2 dotted line). On the other
hand, in the case of $\xi _{R}=5$, when we see current-phase relation $\sin
2\varphi $, we again find two crossing points of ABS (Fig. \ref{fig10}(b)).

\begin{figure}[h]
\begin{center}
\includegraphics[width=60mm]{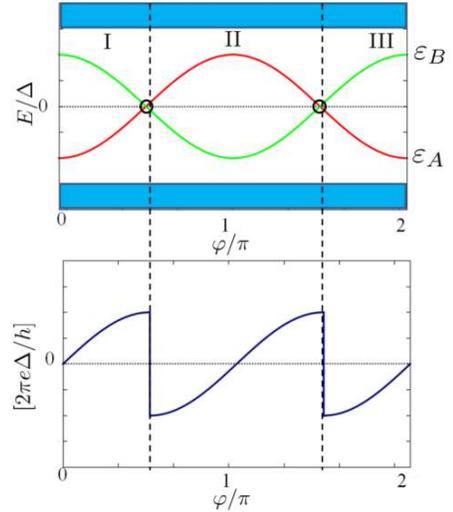}
\end{center}
\caption{(top) A schematic picture of two ABSs denoted by as $\protect%
\varepsilon_A$ (red) and $\protect\varepsilon_B$ (green). These two states
have two crossing points (solid circle) at $E=0$. (bottom) Josephson
current which is given by $J(\protect\varphi)\sim-\frac{\partial\protect%
\varepsilon_A}{\partial\protect\varphi}\text{sgn}(\protect\varepsilon_A)$ }
\label{fig11}
\end{figure}

From these analysis explained above, we find that the behavior of Josephson
current can be determined by the hybridization of ABSs and when this
hybridization produces the two crossing points at the zero energy,
current-phase relation $\sin 2\varphi $ can be seen. Now we consider the
simplified situation where there are two crossing points of ABSs: we have
only two states within the energy gap (blue area in Fig. \ref{fig11} top) which are
labeled by $\varepsilon _{A}$ and $\varepsilon _{B}$ (Fig. \ref{fig11} top). We assume
$\varepsilon _{A}$ ($\varepsilon _{B}$) is written as $\sim -\cos \varphi $ (%
$\cos \varphi $) and crossing points are located at $%
\varphi =\frac{\pi }{2}$ and $\frac{3\pi }{2}$. According to Eq. (\ref{apprx}%
) and particle-hole symmetry, we can focus on the states below the zero
energy to estimate Josephson current. We separate the region of $\varphi $
into three: I $\{\varphi |0\leq \varphi <\frac{\pi }{2}\}$, II $\{\varphi |%
\frac{\pi }{2}\leq \varphi <\frac{3\pi }{2}\}$, III $\{\varphi |\frac{3\pi }{%
2}\leq \varphi <2\pi \}$ (Fig. \ref{fig11} top). In I and III, ABS below the zero
energy obeys $-\cos \varphi $, thus, Josephson current reads $\sim \sin
\varphi $, while in II, ABS transforms into $\sin \varphi $ which gives
Josephson current $-\sin \varphi $. We plot Josephson current as a function
of $\varphi $ ($0\leq \varphi <2\pi $) in bottom of Fig. \ref{fig11}.
Due to the
different curve that ABS obeys in I and II (II and III), there is a
jump at the boundary between I and II (II and III)
which generates current-phase relation $\sin 2\varphi $.
Up to now, $\sin 2\varphi $ dependence of Josepshon current
has been discussed in S/ferromagnetic insulator/S junction,\cite{Tanaka1997}
S/F/S with diffusive F near the
vicinity of $0-\pi$ transition point,
\cite{Golubov_RMP,Sellier,Robinson2007}
$d$-wave superconductor junctions\cite{Yip1993,TK96j,Tanaka97,Josephson3} and
$s$-wave/spin triplet $p$-wave superconductor junctions.\cite{Yip1993,Kwon}
Our setup of realizing $\sin 2\varphi$ dependence is distinct from the preexisting cases.

\subsection{Symmetries of Cooper pair}\label{sec3d}

In this subsection, we focus on the symmetry of Cooper pair in the present
one-dimensional S/F/S junctions. In odd-frequency pairing state, the pair
amplitude changes its sign with the exchange of times of two paired
electrons.\cite{Berezinskii} Taking account of this symmetry class,
symmetry of Cooper pair is classified into (1) even-frequency spin-singlet
even-parity (ESE), (2) even-frequency spin-triplet odd-parity (ETO), (3)
odd-frequency spin-triplet even-parity (OTE), \cite{Berezinskii} and (4)
odd-frequency spin-singlet odd-parity (OSO). \cite{Balatsky} Although
odd-frequency bulk superconductor has not been discovered up to now,
odd-frequency pair amplitude can exist ubiquitously as a subdominant state.
It is known that odd-frequency pairing is induced by the breaking of the
translational \cite{odd3,tanaka12,Eschrig2007} or spin-rotational symmetry \cite{Efetov1,Efetov2,volkov}
from bulk even-frequency pair potential.
Also, it has
been clarified that zero-energy local density of states are enhanced by the
odd-frequency pairing. \cite{odd3,tanaka12,Asano2007PRL,Braude,Asano2007PRB,Eschrig2008}
Odd-frequency pairing influences seriously the proximity effect and various
electronic properties of the junctions.\par

In the present nanowire S/F/S
junctions, the symmetry of pair amplitude far from the S/F (F/S) interface
is ESE $s$-wave one, since the induced pair potential is conventional
spin-singlet $s$-wave. Due to the symmetry breaking, near the S/F (F/S)
interface or inside ferromagnet region, odd-frequency pairings can be induced. Here,
we focus on $s$-wave component of ESE and OTE pair amplitudes in ferromagnet region.
First we focus on the real frequency representation of pair amplitudes. ESE
and OTE amplitudes are given by
\begin{equation}
f_{even}(E)=\frac{1}{2}\bigl\{F_{n,n}^{\uparrow ,\downarrow
}(E+i0^{+})+F_{n,n}^{\uparrow ,\downarrow }(-E-i0^{+})\bigr\}
\end{equation}

\begin{equation}
f_{odd}^{\sigma ,\sigma ^{\prime }}(E)=\frac{1}{2}\bigl\{F_{n,n}^{\sigma
,\sigma ^{\prime }}(E+i0^{+})-F_{n,n}^{\sigma ,\sigma ^{\prime }}(-E-i0^{+})%
\bigr\}
\end{equation}%
with $n\in S_{2}$. $F_{n,n}^{\sigma ,\sigma ^{\prime }}$ comes from Eq. (\ref%
{eq6}):
\begin{equation}
G_{n,n}=\left(
\begin{array}{cc}
G & F \\
\tilde{F} & \tilde{G} \\
\end{array}
\right) ,\;F=\left(
\begin{array}{cc}
F_{n,n}^{\uparrow ,\uparrow } & F_{n,n}^{\uparrow ,\downarrow } \\
F_{n,n}^{\downarrow ,\uparrow } & F_{n,n}^{\downarrow ,\downarrow } \\
\end{array}%
\right)
\end{equation}%
Above representation is useful to compare with the ABS. To clarify the
relation between the ABS and the pair amplitude is an interesting issue,
since it has been revealed that there is a close relation between ABS and
odd-frequency pairing. In the presence of zero energy ABS as a surface state
of unconventional superconductors, odd-frequency pair amplitude is hugely
enhanced.\cite{odd3,odd3b,tanaka12,Bakurskiy} Thus the presence of zero
energy state (ZES) can be interpreted as an emergence of odd-frequency
pairing. Also, it has been clarified that MF always
accompanies odd-frequency pairing. \cite{Asano2013,Wakatsuki,Ebisu,Sau2015}

\begin{figure}[h]
\begin{center}
\includegraphics[width=80mm]{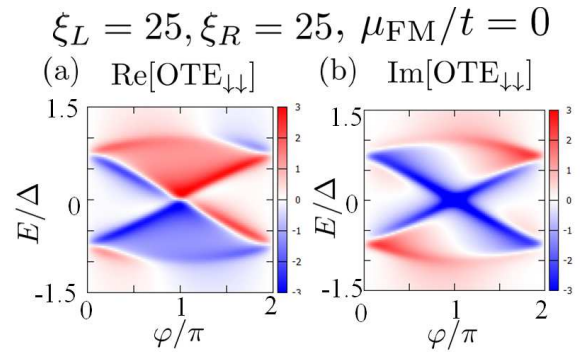}
\end{center}
\caption{(a)(b):The intensity plot of real part (a) and imaginary part (b)
of odd-frequency spin-triplet even-parity (OTE) pair amplitude with $%
\downarrow\downarrow$ spin component $f_{odd}^{\downarrow,\downarrow}(E)$ in
the case of $\protect\xi_L=25$, $\protect\xi_R=25$, $\protect\mu_{\text{FM}%
}/t=0$.}
\label{fig12}
\end{figure}

First, we calculate pair amplitude for $\xi _{L}=25$, $\xi _{R}=25$ and $\mu
_{\text{FM}}/t=0$. The corresponding ABS with the hybridization of Majorana like ZESs
in left and right side has been shown in Fig. \ref{fig7}(a). In this case,
OTE pairing becomes dominant for any phase difference. As compared to other
components, $f_{odd}^{\uparrow ,\uparrow }(E)$, $f_{odd}^{\uparrow
,\downarrow }(E)$ and $f_{even}(E)$, the magnitude of $f_{odd}^{\downarrow
,\downarrow }(E)$ is dominant. This is because that the direction of the
majority spin in ferromagnet is down. Therefore, we plot $s$-wave OTE pairing $%
f_{odd}^{\downarrow ,\downarrow }(E)$ in the middle of the ferromagnet in
Fig. \ref{fig12}. The energy and $\varphi $ dependence of $%
f_{odd}^{\downarrow ,\downarrow }(E)$ is almost similar to those of ABS in
Fig. \ref{fig7}. For $\varphi =0$, $\pi $ and $2\pi $, $\mathrm{Re}%
[f_{odd}^{\downarrow ,\downarrow }(E)]=-\mathrm{Re}[f_{odd}^{\downarrow
,\downarrow }(-E)]$ and $\mathrm{Im}[f_{odd}^{\downarrow ,\downarrow }(E)]=%
\mathrm{Im}[f_{odd}^{\downarrow ,\downarrow }(-E)]$. These relations of
odd-frequency pairing are known in the previous study in normal metal /
superconductor junctions.\cite{odd1}

\begin{figure}[h]
\begin{center}
\includegraphics[width=80mm]{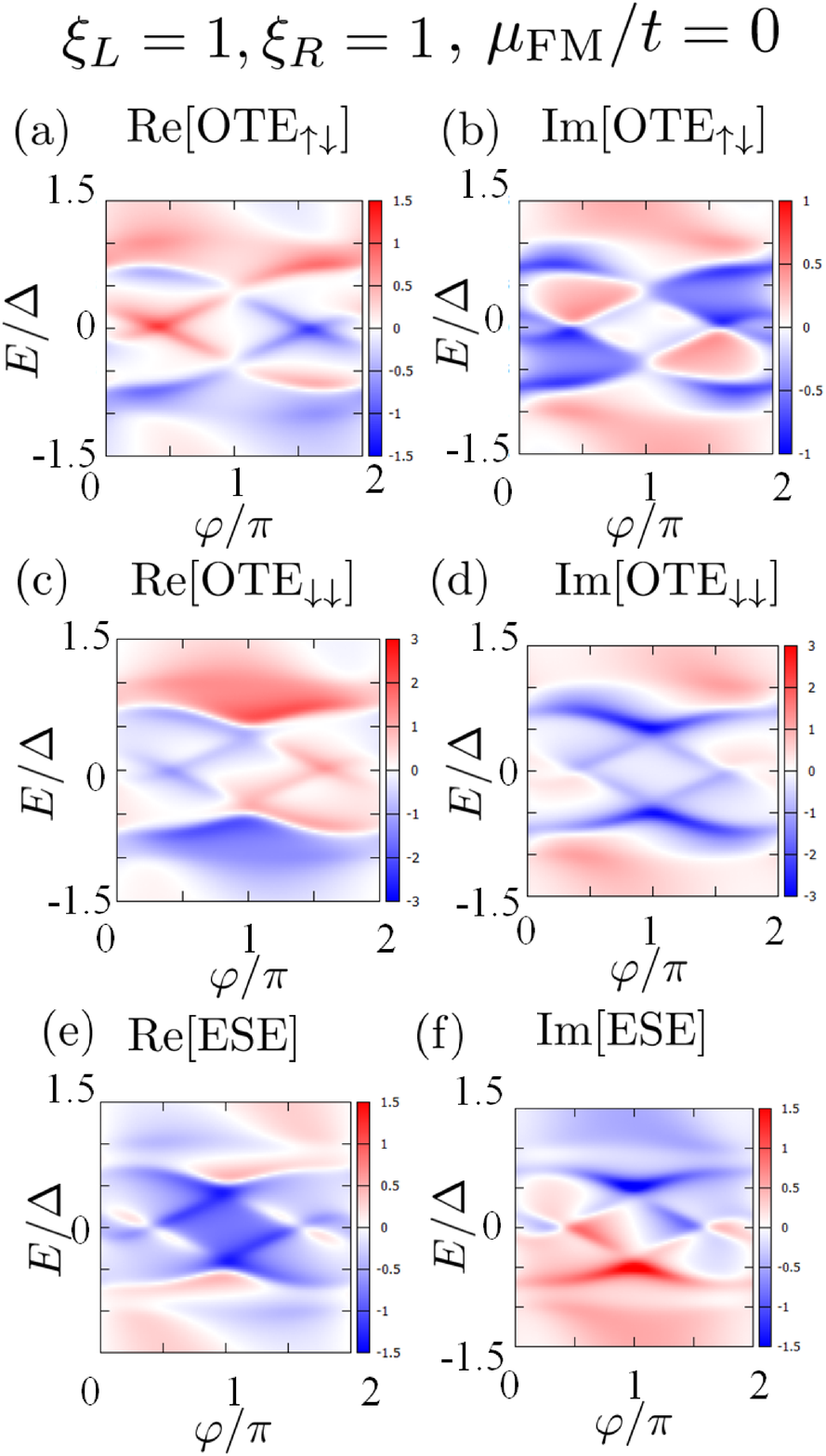}
\end{center}
\caption{(a)-(d): The intensity plot of real part (a)((c)) and
imaginary part (b)((d)) of odd-frequency spin-triplet even-parity (OTE) pair
amplitude with $\downarrow\downarrow$ spin component $f_{odd}^{\downarrow,%
\downarrow}(E)$ and $\downarrow\downarrow$ spin component $%
f_{odd}^{\uparrow,\downarrow}(E)$. In (e) and (f), real part and imaginary
part of even-frequency spin-singlet pair amplitude $f_{even}(E)$ is plotted.
}
\label{fig13}
\end{figure}

Next, we look at the case with $\xi _{L}=\xi _{R}=1$ and $\mu _{FM}=0$. The
corresponding ABS with double crossing points in the energy spectrum of ABS
has been shown in Fig. \ref{fig9}(c). Such characteristic also appears in the
intensity plot of $f_{odd}^{\downarrow ,\downarrow }(E)$, $f_{odd}^{\uparrow
,\downarrow }(E)$ and $f_{even}(E)$. The $E$ and $\varphi $ dependences of $%
f_{odd}^{\uparrow ,\uparrow }(E)$ are similar to those of $%
f_{odd}^{\downarrow ,\downarrow }(E)$. The remarkable point here is that not
only odd-frequency pair amplitude but also even-frequency pair amplitude
exists with the same order in contrast to the case in Fig. \ref{fig13}. For $%
\varphi =0$, $\pi $ and $2\pi $, $\mathrm{Re}[f_{odd}^{\sigma ,\sigma
^{\prime }}(E)]=-\mathrm{Re}[f_{odd}^{\sigma ,\sigma ^{\prime }}(-E)]$ and
$\mathrm{Im}[f_{odd}^{\sigma ,\sigma ^{\prime }}(E)]=\mathrm{Im}%
[f_{odd}^{\sigma ,\sigma ^{\prime }}(-E)]$ with $\sigma =\uparrow
(\downarrow )$ and $\sigma ^{\prime }=\uparrow (\downarrow )$. On the other
hand, $\mathrm{Re}[f_{even}(E)]=\mathrm{Re}[f_{even}(-E)]$ and $\mathrm{%
Im}[f_{odd}(E)]=\mathrm{-{Im}}[f_{odd}(-E)]$ are satisfied.

\section{Summary and Conclusions}\label{sec4}

In this paper, we have studied LDOS, current phase relation of Josephson
current, energy levels of Andreev  bound state and induced odd-frequency pairings in superconductor / ferromagnet / superconductor nanowire junction, where the
magnetization penetrates into superconducting segment with a
decay length $\xi $. We have chosen the chemical potential and SOC so that the topological superconducting regime hosting MF is realized for sufficiently large magnitude of $\xi $. We have found
that when $\xi $ becomes larger, LDOS has a ZEP. On the other hand, if $\xi $ is
shorter, zero energy state at the interface between superconductor and
ferromagnet splits into two states. Accordingly, the behavior of Josephson current
drastically changes. By tuning the parameters of the model, we have found an
almost second-harmonic current-phase relation, $\sin 2\varphi $, with
phase difference $\varphi $. Based on the analysis of ABS, we clarify that current-phase relation is determined by coupling of
the states within the energy gap. We find that the emergence of
crossing points of ABS is a key ingredient to generate $%
\sin 2\varphi $ dependence in current-phase relation. We further studied
both the energy and $\varphi $ dependence of pair amplitudes in ferromagnet region.
For long $\xi $, odd-frequency $s$-wave triplet component is
dominant. The magnitude of the odd-frequency pair amplitude is enhanced at
the energy level of ABS. On the other hand, when $\xi $ becomes shorter, not
only odd-frequency pairing but also even-frequency pairing mixes.\par
Recently, $\sin 2\varphi $
behavior has been observed in S/F/S junctions, \cite{Pal1} when ferromagnet is an insulator which has a spin filter effect. Thus to clarify the
relevance of our obtained $\sin 2\varphi $ dependence to
this experimental report is  a challenging issue.
In our paper, ballistic transport is assumed.
If ferromagnet becomes diffusive,
we can expect anomalous proximity effect
\cite{Proximityp,Proximityp2,Asano2006,Meissner3,Yokoyama2007,Meissner10} by odd-frequency spin-triplet $s$-wave pairing.
Extension to this direction is also an interesting
future study.

\section*{Acknowledgements}
We thank J. Klinovaja, K. T. Law, and S. Kawabata for fruitful discussion.
This work has been supported by Topological Materials Science (TMS) (No. 15H05853),
No. 15H03686, No. 25287085 and No. 15K13498 from the Ministry of Education, Culture, Sports, Science, and
Technology, Japan (MEXT); the Core Research for
Evolutional Science and Technology (CREST) of the Japan
Science and Technology Corporation (JST); Japan-RFBR
JSPS Bilateral Joint Research Projects/Seminars;
Dutch FOM  and the Ministry of Education and Science
of the Russian Federation, grant 14.Y26.31.0007.
H.E. is supported by Grants-in-Aid for JPSP fellow and thanks K.Kawai for useful information.

\appendix
\section{Spatial profile of LDOS}\label{ap1}
In this appendix, we show spatial profile of LDOS of superconductor with decaying ferromagnetic order
to understand ``zero mode-non-zero mode crossover" more clearly. The ferromagnetic order decays from right to left similarly to Fig. \ref{fig3}(a). Since we will consider hybridization of two ZESs on the both edges of superconductor, we set the number of sites rather short, 200. \par

\begin{figure}[h]
\begin{center}
\includegraphics[width=80mm]{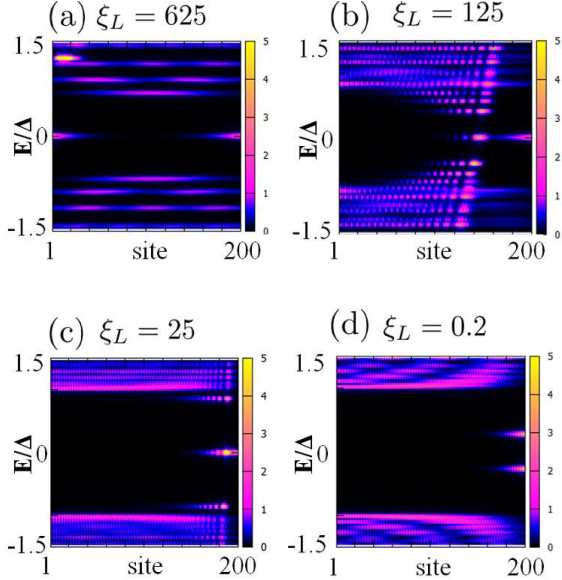}
\end{center}
\caption{(a)-(d) The intensity plot of LDOS of the nanowire with decaying ferromagnetic order on top of the superconductor. The ferromagnetic order decays from right to left similarly to Fig. \ref{fig3}(a).
Horizontal axis represents site index and vertical one does energy. Decay length of the ferromagnetic order is set as (a)$\xi_L=625$, (b)$\xi_L=125$, (c)$\xi_L=25$, and (d)$\xi_L=0.2$.}
\label{dos}
\end{figure}

These profiles are shown in Fig. \ref{dos}. For long enough decay length (Fig. \ref{dos}(a)), ZEPs are localized on the both side of the nanowire, which is analogous to the situation where MFs are localized on the both edges of topological superconductor.\cite{MFpolarizaion} If the decay length is decreased, however, the zero energy mode on the left comes close to the that on the right edge (Fig. \ref{dos}(b)), then finally these two zero modes hybridize (Fig. \ref{dos}(c)) to split into two (Fig. \ref{dos}(d)).
\begin{figure}[h]
\begin{center}
\includegraphics[width=80mm]{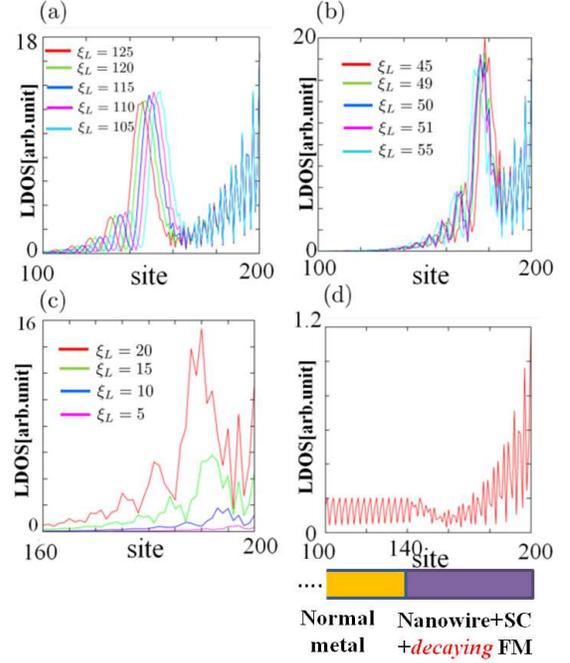}
\end{center}
\caption{(a)-(c) The spatial dependence of LDOS at zero energy of the semiconducting nanowire with pair potential and decaying ferromagnetic order whose length is 200 sites. The several cases of LDOS are plotted for different decay length which is indicated by different colors as shown in each figure. (d) The spatial dependence of LDOS of the normal metal/nanowire junction system. The total length of this junction is set as 200 sites and the interface between normal metal and nanowire is located at site 140.
In the nanowire segment, the pair potential and decaying ferromagnetic order with $\xi_L=125$ are also included similarly to (a)-(c). A schematic picture of this junction is shown below the plot.}
\label{apfig1}
\end{figure}
\begin{figure}[h]
\begin{center}
\includegraphics[width=80mm]{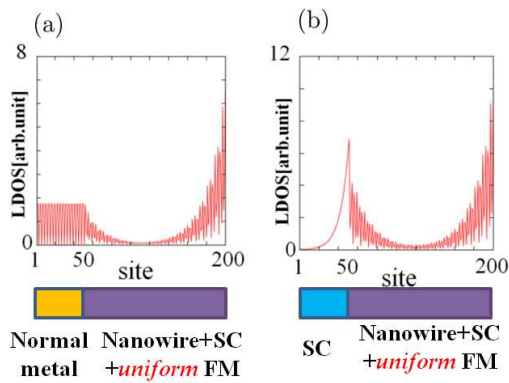}
\end{center}
\caption{(a) The spatial dependence of LDOS of the normal metal/nanowire with pair potential and {\it uniform} ferromagnetic order junction system whose length is 200 sites. The interface is positioned at 50 sites. (b) The similar plot to (a) but the normal metal is replaced by $s$-wave superconductor. Schematic pictures of the junction are shown below each plot.
}
\label{apfig2}
\end{figure}

Below, we analyze the spatial profile of LDOS of the system we consider here in more detail especially focusing on at the zero energy. In Figs. \ref{apfig1}(a)-(c), we plot this spatial dependence of LDOS for several cases of decay length. For large enough $\xi_L$ (Fig. \ref{apfig1}(a) and (b)), we see the two ZEP: one is on the edge, $i.e.$ at 200 sites and another which is identified as broad peak is left from it (for Fig. \ref{apfig1}(a), this ZEP is located at site $140\sim150$ and for Fig. \ref{apfig1}(b), it is at around site $180$). We notice that the position of the left peak is shifted if the decay length is changed and accordingly the behavior of the oscillations of LDOS left from the broad peak changes. However, we also find that the oscillations between the two peaks show the similar behavior. When the decay length is shorter than around $\xi_L=20$, the two peaks hybridize and are away from the zero energy (Fig. \ref{apfig1}(c)). Further, we also calculate the spatial dependence of LDOS of normal metal/nanowire junction system with $\xi_L=125$. We set the interface at the site 140. The result is shown in Fig. \ref{apfig1}(d). The board peak which is indicated in Fig. \ref{apfig1}(a) spreads into the normal metal.\par Actually, we can regard the two ZEPs explained above as the MFs. To make this statement more convincing, we calculate the spatial profile of the LDOS of the normal metal/nanowire with pair potential and {\it uniform} ferromagnetic order junction system as well as that of $s$-wave superconductor/nanowire junction system. These plots are shown in Figs.  \ref{apfig2}(a) and (b), respectively. The interface is located at site 50. In Fig. \ref{apfig1}(a), the ZEP on the left which should be expected to appear at the interface penetrates into the normal metal segment. This is analogous to Fig. \ref{apfig1}(d). Moreover, in Fig. \ref{apfig2}(b), we can see the ZEP on the left (at site 50), however, as opposed to Figs. \ref{apfig2}(a), this ZEP cannot spread into the left from the interface due to the superconducting energy gap. Therefore, ZEP cannot be stabilized and decays away from the interface. This behavior is similar to Figs. \ref{apfig1}(a)(b), however, there is one difference: in Figs. \ref{apfig1}(a)(b) due to the presence of {\it non-uniform} ferromagnetic order, there isn't the explicit boundary which distinguishes the topologically trivial area and non-trivial one. Thus, the broad ZEP appears in Figs. \ref{apfig1}(a)(b). \par
The comparison between the results of Fig. \ref{apfig1} and Fig. \ref{apfig2} implies that two ZEPs appeared in {\it non-uniform} case can be identified as two MFs.
\bibliography{nanowire1}
\end{document}